\documentclass[12pt]{article}

\usepackage{epsfig}

\usepackage{amssymb}
\usepackage{amsfonts}

\usepackage{color}
\def\be{\begin{equation}}
\def\ee{\end{equation}}
\def\ba{\begin{array}{c}}
\def\ea{\end{array}}

\newcommand{\bea}{\begin{eqnarray}}
\newcommand{\eea}{\end{eqnarray}}

\newcommand{\bbr}{\br\!\br}
\newcommand{\kkt}{\kt\!\kt}

\newcommand{\pkt}{\!\succ\,\,}
\newcommand{\kt}{\rangle}
\newcommand{\br}{\langle}

\begin{document}

\begin{center}

{\Large \bf

Discrete-coordinate crypto-Hermitian quantum system controlled by
time-dependent Robin boundary conditions

}

\end{center}

\vspace{0.8cm}

\begin{center}

  {\bf Miloslav Znojil}$^{a,b,c}$

\end{center}

 $^{a}$
The Czech Academy of Sciences,
 Nuclear Physics Institute,
 Hlavn\'{\i} 130,
250 68 \v{R}e\v{z},

Czech Republic,
 {e-mail: znojil@ujf.cas.cz}

 $^{b}$  Department of Physics, Faculty of
Science, University of Hradec Kr\'{a}lov\'{e}, Rokitansk\'{e}ho 62,

50003 Hradec Kr\'{a}lov\'{e},
 Czech Republic,
 {e-mail: miloslav.znojil@uhk.cz}

 $^{c}$
Institute of System Science, Durban University of Technology,
Durban, South Africa

\vspace{10mm}

\subsection*{Abstract}

A family of exactly solvable quantum square wells with discrete
coordinates and with {certain} {\em non-stationary\,}
Hermiticity-violating Robin boundary conditions is proposed and
studied. Manifest non-Hermiticity of the model in
conventional Hilbert space ${\cal H}_{friendly}$ is required to
coexist with the unitarity
{of}
system in another, {\it ad hoc\,}
Hilbert space ${\cal H}_{physical}$. Thus,
quantum mechanics in its non-Hermitian interaction picture (NIP)
{representation is to be used}.
We must construct the time-dependent states (say,
$\psi(t)$) as well as the time-dependent observables (say,
$\Lambda(t)$). Their evolution {in time
is} generated by
the operators denoted,
here, by {the respective}
symbols $G(t)$ (a Schr\"{o}dinger-equation generator)
and $\Sigma(t)$ (a Heisenberg-equation generator, a.k.a.
quantum Coriolis force). The unitarity {of evolution}
in ${\cal
H}_{physical}$ is then guaranteed by the reality
{of}
spectrum of
the energy observable {\it alias\,} Hamiltonian
$H_{}(t)=G_{}(t)+\Sigma_{}(t)$. {The applicability
of these ideas is illustrated via} an
$N$ by $N$ matrix model. At $N=2$, closed
formulae are presented not only for the measurable
instantaneous energy spectrum but also for all of the eligible
time-dependent physical inner-product metrics $\Theta_{(N=2)}(t)$,
for the related Dyson maps $\Omega_{(N=2)}(t)$, for the Coriolis
force $\Sigma_{(N=2)}(t)$ as well as, in the very ultimate step of
the construction, for the truly nontrivial Schr\"{o}dinger-equation
generator $G_{(N=2)}(t)$.

\subsection*{Keywords}

quantum theory of unitary systems; non-Hermitian interaction
representation; non-stationary physical inner products; solvable
discrete square well;

\newpage

\section{Introduction\label{sI}}

Among applications of quantum mechanics working with
{stationary} observables which are non-Hermitian but
quasi-Hermitian \cite{Geyer}, a comparatively exceptional position
is {taken} by the theories in which the information about dynamics
is merely carried by boundary conditions
\cite{Matra,Matrb,Matrc,Fermi}. One of the {simplest,}
square-well
examples of such a type can be found discussed in \cite{Coronado}.
In the context of scattering
the boundary conditions of a manifestly non-Hermitian form have been
assigned there a fully conventional physical interpretation of a
perfect-transmission constraint.
A less artificial-looking and, currently, {more}
widely accepted physical  {bound-state}
treatment of all of the analogous non-Hermitian
but quasi-Hermitian (i.e., hiddenly Hermitian) models and
Hamiltonians can be found described in a number of reviews
{in which
one finds the
very general mathematical
concept of quasi-Hermitian operators \cite{Dieudonne}
narrowed
to the operators which are quasi-Hermitian but bounded \cite{Geyer}
or merely $\eta-$pseudo-Hermitian {\it alias\,}
pseudo-Hermitian \cite{ali}
or even just
parity-pseudo-Hermitian {\it alias\,}
parity-time symmetric \cite{Carl}}.

In the formulation
of {stationary} quantum theory
called non-Hermitian
Schr\"{o}dinger picture (NSP) one
has to {combine} physics ({requiring} the
unitarity
{of}
evolution which is just guaranteed in
a certain {hypothetical,}
user-unfriendly Hilbert space ${\cal H}_{physical}$)
{with} mathematics
({set in another space ${\cal H}_{mathematical}$ and}
needed during the practical
implementation of the theory).

In the conventional {stationary}
quantum mechanics of textbooks \cite{Messiah},
both of the latter two Hilbert spaces coincide. In the upgraded NSP
version of the theory (in which one admits that
${\cal H}_{physical} \neq {\cal H}_{unphysical}$, cf. also several
more rigorous and updated reviews in \cite{book}) one always has to
keep in mind that any operator $\Lambda$ representing an observable
is, by definition, simultaneously self-adjoint in the hypothetical
``correct space'' ${\cal H}_{physical}$ and non-Hermitian in the
friendlier, ``unphysical'' but preferred construction space ${\cal
H}_{mathematical}$.

In practice it appeared convenient to work just in ${\cal
H}_{unphysical}={\cal H}$. Using the common notation
convention one stays in ${\cal H}$, writes $\Lambda\neq
\Lambda^\dagger$ and introduces the so called inner-product metric
$\Theta$ in order to guarantee the observability status of $\Lambda$
via the Dieudonn\'e's \cite{Dieudonne} quasi-Hermiticity postulate
$\Lambda^\dagger\Theta=\Theta\,\Lambda$ (cf. also \cite{Geyer} for
details). In such a setting one treats ${\cal H}_{physical}$
as represented in ${\cal H}_{unphysical}$ while one only amends
the inner product,
 $$
 \br \psi_1|\psi_2\kt_{physical}=\br
 \psi_1|\Theta|\psi_2\kt_{unphysical}\,.
 $$
In the second item the subscript $_{unphysical}$ can and will
be omitted as superfluous.

During the early years of development and applications of the NSP
formalism people believed
that for unitary systems (of our present
interest) ``the inner product of the physical Hilbert space cannot
depend on time unless one defines the dynamics of the quantum
system by an operator that is not observable''
{(cf. Theorem 2 and
subsequent comments on p. 1272 in \cite{ali})}.

Later, it became
clear that the applicability of the latter ``no-go'' Theorem is
restricted just to the mere NSP framework in which, generically, the
non-Hermitian observables remain stationary, time-independent,
$\Lambda^{(NSP)} \neq \Lambda^{(NSP)}(t)$.
An ultimate remedy and
clarification of the misunderstanding has been found, in 2008, in
{a non-stationary}
extension of the
{quasi-Hermitian formulation of}
quantum mechanics
{(to be
called, in what follows,
non-Hermitian interaction picture, NIP, cf. its introduction in
\cite{timedep,SIGMA})}.

{A few years later,
the situation has been re-analyzed \cite{Gong}
and
the adequacy of the
use of time-dependent metric operators
has been reconfirmed \cite{Fring}
(in this respect cf. also the recent comprehensive NIP
review \cite{NIP}).}
In the NIP framework of our present interest (where one
admits the non-stationarity $\Theta_{NIP} = \Theta_{NIP}(t)$) the
necessary mathematics has been found perceivably more
complicated. For this reason, in a way inspired by the recent NSP
description of the role
{of}
non-Hermitian but stationary boundary
conditions in a discrete Schr\"{o}dinger equation \cite{2014} we are
now going to describe a discrete but still unitary quantum system in
which the boundary conditions would be not only non-Hermitian but
also non-stationary.

A detailed formulation of the problem  as well as a constructive
sample of its solution will be given. We will outline the basic
features of the properly amended NIP alternative to the more
traditional stationary NSP formulation of quantum theory. After a
concise exposition
{of}
necessary mathematics we will redirect
emphasis to physics. We will explain how the ``input'' knowledge of
the dynamics-determining non-Hermitian and time-dependent boundary
conditions can be consequently converted into a consistent
theoretical scheme yielding the ``output'' predictions of the
results of measurements
{of}
observable characteristics
{of a}
non-stationary but stable{,
hiddenly} unitary physical
system.

The presentation of our results will be arranged as follows. First,
in section \ref{sII} we will outline the basic features
{of the}
formalism. Then, section \ref{sIII} will
be devoted to the introduction of our specific boundary-interaction
model. This will be followed by section \ref{s4I} in which our
attention will be turned to the construction and properties of the
physical Hilbert space.
{In}
subsequent section~\ref{sIIIc2} we
will display the explicit illustrative formulae while in our final
sections \ref{sIVdi} and \ref{sIr} we will add a few final remarks
and conclusions.

\section{Hiddenly Hermitian quantum mechanics {\it in nuce}\label{sII}}

\subsection{Stationary cases and NSP physical Hilbert spaces}

Given an arbitrary non-Hermitian
Hamiltonian
{$H \neq H^\dagger$},
the first question to ask concerns the consistent
probabilistic interpretation of the model. {The}
answer was
formulated, in 1992, by Scholtz et al \cite{Geyer}. These authors
explained how the conventional requirement
{of}
self-adjointness
of a Hamiltonian can be weakened.
{In particular, they} emphasized that in the
analysis of many realistic systems the uniqueness of a conventional
textbook Hilbert space of states (say, ${\cal L}$, which has to be,
{\em simultaneously}, user-friendly and physical \cite{Messiah}) may
happen to be over-restrictive.

They proposed to split {the} roles and
to work, simultaneously, with the two separate, non-equivalent
Hilbert spaces. Naturally, a preselected Hamiltonian $H$ (or any
other observable $\Lambda$) can only be self-adjoint in one of them
(i.e., say, in ${\cal H}_{physical}$). The other, non-equivalent
Hilbert space (i.e., say,
{${\cal H}_{mathematical}={\cal H}_{friendly}$})
may be preferred,
as {\em the\,} representation space {in calculations.}
At the same time, {the latter space}
must necessarily be
perceived as
manifestly unphysical since
{in this space one has $H \neq H^\dagger$}.

For our present purposes {the}
two Hilbert spaces may be
interpreted as complementary since ${\cal H}_{friendly}={\cal
H}_{unphysical}$ and ${\cal H}_{physical}={\cal H}_{unfriendly}$.
The main motivation of the split is that for many quantum systems
of practical interest the innovated formulation of quantum mechanics
might be more calculation-friendly. Moreover, it was of paramount
importance to imagine that the correct physical Hilbert
space ${\cal H}_{unfriendly}$ appeared to be comparatively easily
represented in the mathematically more suitable working space ${\cal
H}_{friendly}$ \cite{Geyer}. What appeared sufficient was the mere
amendment
{of}
inner product (the related technical details may be
also found discussed in reviews \cite{ali,Carl,book}).

In applications (and, in particular, in applications in which the
operators
{of}
observables are stationary, time-independent),
it is very natural to expect that
{all of the necessary calculations will be simpler}
after transition
from the conventional Schr\"{o}dinger equation living
in ${\cal L}$,
viz., from {the equation of textbooks}
 \be
 {\rm i}\,\frac{d}{dt}\,|\psi^{}(t)\pkt
 =
 \mathfrak{h}\,|\psi^{}(t)\pkt\,,
 \ \ \ \ |\psi^{}(t)\pkt \in {\cal L}
 \,,
 \ \ \ \ \mathfrak{h} = \mathfrak{h}^\dagger
 \label{pCauchy}
 \ee
to its alternative
living in ${\cal H}_{friendly}$,
 \be
 {\rm i}\,\frac{d}{dt}\,|\psi^{}(t)\kt
 =
 H\,|\psi^{}(t)\kt\,,
 \ \ \ \ |\psi^{}(t)\kt \in {\cal H}_{friendly}\,,
 \ \ \ \ H \neq H^\dagger\,.
 \label{SCauchy}
 \ee
The price to
pay for the change is that the new version
{of}
Hamiltonian (which is self-adjoint in ${\cal
H}_{unfriendly}$) appears non-Hermitian in the
mathematical representation space{.}

\subsection{Non-stationary NIP and physical Hilbert space}

In the
{NSP} formulation of quantum mechanics
one of the most essential assumption is that
the
one-to-one correspondence
between the Hilbert spaces and/or between the Schr\"{o}dinger
equations
(realized, say, by an invertible ``Dyson map'' operator
$\Omega$ \cite{Dyson})
remains time-independent
(cf., e.g., Theorem Nr. 2 in \cite{ali}).
After a more or less straightforward
non-stationary NIP generalization of the theory,
{unfortunately,}
only too many changes did occur.

The first one was that
in the specification
{of}
correspondence
$|\psi^{}(t)\pkt \leftrightarrow |\psi^{}(t)\kt$
we had to keep in mind, in general, the manifest
time-dependence of all of the relevant operators. Thus,
using a time-dependent generalization $\Omega=\Omega(t)$
of the invertible Dyson-map
operator we postulate
 \be
 |\psi^{}(t)\pkt
 =\Omega(t)\,|\psi^{}(t)\kt\,,
 \ \ \ \ \
 \mathfrak{h}(t)=
 \Omega(t)\,H(t)\,\Omega^{-1}(t)\,.
 \label{[7]}
 \ee
{Due}
to the emergence of
a non-vanishing Coriolis-force operator
 \be
 \Sigma(t)={\rm i}\,\Omega^{-1}(t)\,\dot{\Omega}(t)
 \label{7b}
 \ee
(where the dot over ${\Omega}(t)$ marks the differentiation with respect to time)
the insertion of the textbook ket
$|\psi^{}(t)\pkt =\Omega(t)\,|\psi^{}(t)\kt$
does not convert the textbook Schr\"{o}dinger Eq.~(\ref{pCauchy})
into its NSP partner (\ref{SCauchy}) but rather into its modified, NIP partner
 \be
 {\rm i}\,\frac{d}{dt}\,|\psi^{}(t)\kt
 =
 G(t)\,|\psi^{}(t)\kt\,,
 \ \ \ \ |\psi^{}(t)\kt \in {\cal H}_{friendly}\,,
 \ \ \ \ G(t)=H(t)-\Sigma(t) \,.
 \label{DCauchy}
 \ee
In contrast to the
stationary NSP models with vanishing $\Sigma(t)=0$,
we now have, in general, $G(t)\neq H(t)$.
This means that the ``dynamical information input''
knowledge of the textbook
Hamiltonian $\mathfrak{h}(t)$
(or, more precisely, of its
non-Hermitian isospectral image $H(t)$ defined as acting
in ${\cal H}_{friendly}$)
does not still enable us to
write down Schr\"{o}dinger Eq.~(\ref{DCauchy}) and, via its solution,
to reconstruct the evolution
{of ket-vectors}
$|\psi^{}(t)\kt \in {\cal H}_{friendly}$.

Fortunately, a hypothetical knowledge of the
time-dependence
{of}
mapping  ${\Omega}(t)$
and of the inner-product metric
$\Theta(t) = \Omega^\dagger(t)\,\Omega(t)$
{enables}
us to
re-express the self-adjointness
{of}
$\mathfrak{h}(t)$ in ${\cal L}$
via the time-dependent generalization
of the
Dieudonn\'{e}'s quasi-Hermiticity property
of
its isospectral observable-Hamiltonian avatar
$H(t)=\Omega^{-1}(t)\,\mathfrak{h}(t)\, \Omega(t)$
in ${\cal H}_{friendly}$
\cite{Dieudonne},
 \be
 H^\dagger(t)\,\Theta(t)=\Theta(t)\,H(t)\,.
 \label{tJD}
 \ee
Hence, we may invert the flowchart
{and}
assume that given the
Hamiltonian (with real spectrum) in its non-Hermitian-representation
version $H(t)$ (preselected and defined as acting in ${\cal H}_{friendly}$),
the necessary search
{for}
inner-product metric $\Theta(t)$
can still be based on the solution of~(\ref{tJD}), i.e., of
the {linear}
equation.
We may conclude that in the non-stationary NIP framework the
construction
{of}
metric
{ $\Theta(t)$}
can proceed in full
analogy with the NSP recipes{.}

After the construction of $\Theta(t)$ one is also allowed to employ
the standard operator methods and to construct the ``square root''
$\Omega(t)$
{of}
metric as well as
{its}
inverse
and the time
{derivative}
as needed in
Eq.~(\ref{7b}). The necessity
{of}
construction of all of these
``missing'' components
{of}
Coriolis force represents, in fact, a
new and difficult technical challenge. Only its satisfactory
resolution may enable us to define, ultimately, the NIP
Schr\"{o}dinger equation and to construct the ket vectors
representing the states (cf. also review \cite{NIP} for details).

Only on this basis we would finally be able to
restore the NIP-NSP parallels and to
predict the
results
{of}
measurements.
Typically, whenever one considers an observable of interest
(represented by an operator $\Lambda(t)$ with real spectrum and such that
$\Lambda^\dagger(t)\,\Theta(t)=\Theta(t)\,
\Lambda(t)$),
the NIP predictions
will be based again on
the evaluation
{of}
{overlaps}
 \be
 \br \psi(t)|\,\Theta(t)\,\Lambda(t)|\psi(t)\kt\,.
 \label{eme}
 \ee
{The} presence of the correct physical
inner-product metric $\Theta(t)\neq I$ indicates that
these overlaps
only
have their correct probabilistic interpretation
in
the
truly anomalous and, {through}
non-stationary metric, manifestly
{time-dependent form of}
Hilbert space
{${\cal H}_{physical}={\cal H}_{physical}(t)$}.

\section{Boundary-controlled square-well model\label{sIII}}

The existence
{of}
NIP-related ``new and difficult technical
challenges'' as mentioned in preceding section was one of the main
sources of inspiration of our forthcoming detailed and constructive
study of the non-stationary version of the discrete square-well
model endowed with nontrivial, {manifestly
time-dependent}
boundary conditions.

{Before we start addressing the related technical
challenges, let us briefly mention that
in a broader physical context,
the studies
{of}
quantum systems characterized
{by}
time-dependent
boundary conditions are currently finding phenomenological
applications
which range from
mathematical and condensed-matter physics to cosmology
(for a concise reference
let us just cite the recent preprint \cite{Jarda}).
In this setting,}
our choice
{of}
model
 \be
 H(t)=
 \left[ \begin {array}{ccccc}
  2-z(t)&-1&0
 &\ldots&0
 \\
 \noalign{\medskip}-1&2&-1&\ddots&\vdots
 \\
 \noalign{\medskip}0&-1&\ddots&\ddots
 &0
 \\
 \noalign{\medskip}\vdots&\ddots&\ddots&2&-1
 \\
 \noalign{\medskip}0&\ldots&0&-1&2- z^*(t)
 \end {array} \right]\,
 \label{Ka8t}
 \ee
was strongly encouraged by the dynamical input knowledge
status of the {boundary-value
matrix elements $z(t) \in \mathbb{C}$}
of {our potentially}
observable time-dependent {quantum}
Hamiltonian.

Our mathematically motivated choice of such an $N$ by $N$ matrix
model with $N < \infty$ will also decisively facilitate the
physics-oriented construction of the predictions
{of}
measurements
(\ref{eme}).

\subsection{Robin boundary conditions\label{s2I}}

Many salient
features {of}
bound-state problems of conventional
textbooks \cite{Messiah} {are}
well illustrated by the exactly solvable ordinary differential
square-well Schr\"{o}dinger equation \cite{Fluegge}
 \be
 -\frac{d^2}{dx^2}\psi_n(x)=\varepsilon_n\psi_n(x)
  \,,\ \ \
   \psi_n(0)= \psi_n(L)=0\,
 \label{spojhambc}
 \ee
and/or by its numerically motivated difference-equation
equidistant-lattice analogue \cite{Acton}
 \be
 -\psi_n(x_{k-1})+2\,\psi_n(x_{k})-\psi_n(x_{k+1})=
 E_n^{(N)}\psi_n(x_{k})
 \,,\ \ \
 \psi_n(x_{0})=\psi_n(x_{N+1})=0\,
 \label{diskrhambc}
 \ee
where $ n = 0, 1, \ldots$ and where either $x \in (0,L)$ or $ k = 1,
2,\ldots,N$, respectively.
One of the important new methodical merits of both of the latter two
old toy models emerged, recently, in the framework of the so called
non-Hermitian reformulations of quantum mechanics: For our present
purposes we may recall, in this respect, either the older reviews
\cite{Geyer,ali,Carl} (and speak about a stationary non-Hermitian
Schr\"{o}dinger picture, NSP) or paper \cite{timedep} and newer
reviews \cite{book,NIP} (and speak about a non-stationary
non-Hermitian interaction picture, NIP).

{We are returning}
to these questions with a new motivation
provided
{by}
the emergence of
{difficulties} accompanying
the growth of interest in certain {\em
non-stationary\,} NIP models \cite{Fring,Mousse}. During the
formulation of our present project we felt encouraged by the mutual
relationship between the two square-well Schr\"{o}dinger equations
(\ref{spojhambc}) and {(\ref{diskrhambc})}.
In parallel, a
strictly phenomenological source of our interest can be seen in a
consequent restriction of the ``input'' information about dynamics
to the boundaries, i.e., in a certain ``minimality'' of the
non-Hermitian ingredients in these models.

{We}
decided to replace the conventional Dirichlet boundary conditions by
their Hermiticity-violating two-parametric (i.e.,
Robin-boundary-condition) alternatives
 \be
  \psi(0)= \frac{\rm i }{\alpha + {\rm i}\beta}
  \,\frac{d}{dx}\psi(0)\,,\ \ \ \ \ \
  \psi(L)= \frac{\rm i }{\alpha - {\rm i}\beta}
  \,\frac{d}{dx}\psi(L)\,
   \label{bcdavid}
 \ee
(in (\ref{spojhambc}), with two free real parameters $\alpha \,,\,
\beta\, \in \,\mathbb{R}$) or
 \be
  \psi_n(x_{0})= \frac{\rm i }{\alpha + {\rm i}\beta}\,
  \left (
  \frac{\psi_n(x_{1})-\psi_n(x_{0})}{h}
  \right )\,,
  \ \ \ \ \ \
  \psi_n(x_{N+1})= \frac{\rm i }{\alpha - {\rm i}\beta}\,
  \left (
  \frac{\psi_n(x_{N+1})-\psi_n(x_{N})}{h}
  \right )\,
 \label{prvnice}
  \ee
(in (\ref{diskrhambc}), with a suitable lattice grid-point distance
$h>0$), respectively.

{A}
wealth of consequences may be expected to emerge. The
most {obvious}
one lies in the necessity of an upgrade of the
conventional formulation of quantum mechanics.
{A}
key
challenge emerges due to our innovated interpretation
{of}
parameters in conditions (\ref{bcdavid}) and (\ref{prvnice}) which
will be allowed non-stationary, time-dependent,
 \be
 \alpha=\alpha(t)\,,\ \ \ \ \beta=\beta(t)\,.
 \ee
The latter, innocent-looking generalization leads to a number of
nontrivial constructive tasks. In the forthcoming{,}
methodically
sufficiently instructive analysis only the difference
Schr\"{o}dinger-equation model will be considered.

 \subsection{Condition number one: the reality of spectrum}

The two stationary {and}
manifestly non-Hermitian square-well bound-state problems
as mentioned in Introduction were thoroughly studied,
in the NSP framework, in \cite{2014}. We noticed there that the
differential-equation problem
can be {perceived}
as a specific
(i.e., $h \to 0$ and $N \to \infty$) special-case limit of its
difference-equation partner. Thus, we just
studied
the
difference-equation bound-state problem with finite $N$
and with a single complex parameter $z=1/(1 - \beta\,h
-i\alpha\,h)$.

Once we rewrote the corresponding stationary Schr\"{o}dinger equation
in its equivalent $N$ by $N$ matrix form in ${\cal H}_{friendly}$,
 \be
 \left[ \begin {array}{ccccc}
  2-z&-1&0
 &\ldots&0
 \\
 \noalign{\medskip}-1&2&-1&\ddots&\vdots
 \\
 \noalign{\medskip}0&-1&\ddots&\ddots
 &0
 \\
 \noalign{\medskip}\vdots&\ddots&\ddots&2&-1
 \\
 \noalign{\medskip}0&\ldots&0&-1&2- z^*
 \end {array} \right]\,
  \left[ \begin {array}{c}
 \phi_1\\
 \noalign{\medskip}\phi_2\\
 \noalign{\medskip}\vdots\\
 \noalign{\medskip}\phi_{N-1}\\
 \noalign{\medskip}\phi_N\\
 \end {array} \right]
 =
 E_n^{(N)}\,
  \left[ \begin {array}{c}
 \phi_1\\
 \noalign{\medskip}\phi_2\\
 \noalign{\medskip}\vdots\\
 \noalign{\medskip}\phi_{N-1}\\
 \noalign{\medskip}\phi_N\\
 \end {array} \right]
 \,
 \label{Ka8p}
 \ee
(cf. Eq. Nr. 13 in \cite{2014}) it was comparatively straightforward
to reveal the exact solvability of the problem.
Indeed, after abbreviation $2-E_n^{(N)} =2y$
our
Schr\"{o}dinger Eq.~(\ref{Ka8p}) acquired, strictly, the form
{of}
recurrences satisfied
{by}
Chebyshev polynomials of the first and second kind
\cite{Ryshik}.
In other words,
we could set
 \be
 \phi_{n}=A\,T_{n-1}(y)+B\,U_{n-1}(y)\,,\ \ \ \ n = 1, 2, \ldots,
 N\,.
 \label{ansatz}
 \ee
Subsequently, we could fix the values of the
two complex parameters $A$ and $B$
and of the energy
via the normalization and boundary conditions, i.e.,
via the first and last line of Eq.~(\ref{Ka8p}).

Precisely this has been done in \cite{2014}.
Serendipitously we discovered {there} that the
Hamiltonian in (\ref{Ka8p}) is PT-symmetric,
$HPT=PTH$.
Whenever this symmetry
proves spontaneously unbroken,
the energies are all real,
i.e., the {evolution}
in time remains unitary \cite{Carl}.
We proved that the existence of such a
dynamical regime
is guaranteed in a non-empty complex vicinity of real $z=1$
(cf. Proposition Nr. 1 in {\it loc. cit.}).

The quantum system in question
has also been given the standard probabilistic interpretation.
Realized, in some cases,
via explicit formulae determining a suitable stationary NSP metric $\Theta$
at any $N$
(cf., e.g., Proposition Nr. 2 in {\it loc. cit.}).

 \subsection{Numerics and reparametrizations}

In our present paper we decided to extend the latter analysis
to the non-stationary dynamical regime in which the
complex dynamics-controlling parameter
becomes allowed to vary with time,
$z=z(t)$.
The motivation of such a project was threefold.
Firstly, we felt encouraged by the fact that
the introduction $z \to z(t)$ of the non-stationarity
of dynamics
leaves the formal solvability
{of}
eigenvalue problem (\ref{Ka8p})
via ansatz (\ref{ansatz})
unchanged.
Secondly, we imagined that the
extremely elementary
one-parametric nature of our $N$ by $N$
Hamiltonian $H(t)$
enhances the chances of
the constructive considerations being successful.

Thirdly, having performed a few preliminary tests at $N=2$
we revealed that in a way paralleling a few stationary-theory
observations as made in \cite{2014},
the insight in the structure
and properties
{of}
bound states
can significantly be enhanced
{when}
parameter $z \in \mathbb{C}$
gets replaced by a more specific
real {variable.}
{An amended} insight
emerged when we replaced the value of $z$
by its redefinition $\, i\sqrt {1-{r}^{2}}\,$ using
{a}
real variable $r$
and yielding
 \be
 H=
 \left[ \begin {array}{cc} 2-i\sqrt {1-{r}^{2}}&-1\\\noalign{\medskip}
-1&2+i\sqrt {1-{r}^{2}}\end {array} \right]\,.
\label{prodva}
 \ee
A key merit of such a (still, stationary)
{reduction}
appeared to lie not only in the extremely
elementary form
{of}
spectrum $E^{(2)}_\pm = 2 \pm r$
but also in
{a}
serendipitous discovery that the matrix $H$
ceases to be diagonalizable in the limit of
{vanishing} $r \to 0$.

In this limit, in the language of mathematics, the model acquires
the Kato's \cite{Kato}
exceptional-point (EP) singularity.
Hence, the usual diagonal-matrix
representation
{of}
Hamiltonian
 \be
  \mathfrak{h}_S=\mathfrak{h}_S(r)=
  \left[ \begin {array}{cc} r+2&0\\\noalign{\medskip}0&-r+2\end {array}
 \right]\,
\label{17S}
 \ee
remains restricted to $r\neq r^{(EP)}=0$. Alternatively,
the latter matrix becomes
tractable also as the special {diagonal} self-adjoint
textbook Hamiltonian in ${\cal L}$.

Even in the language of physics,
the choice of $r=0$ must be excluded as not compatible with the
postulates of consistent quantum theory of closed systems.
At EP, matrix (\ref{17S})
has to be
replaced
by a canonical non-diagonal Jordan block
$$
  \mathfrak{h}_S(0)=
  \left[ \begin {array}{cc} 2&1\\\noalign{\medskip}0&2\end {array}
 \right]\,.
$$
in a process which is discontinuous in $r$.

\begin{figure}[h]
\begin{center}
\epsfig{file=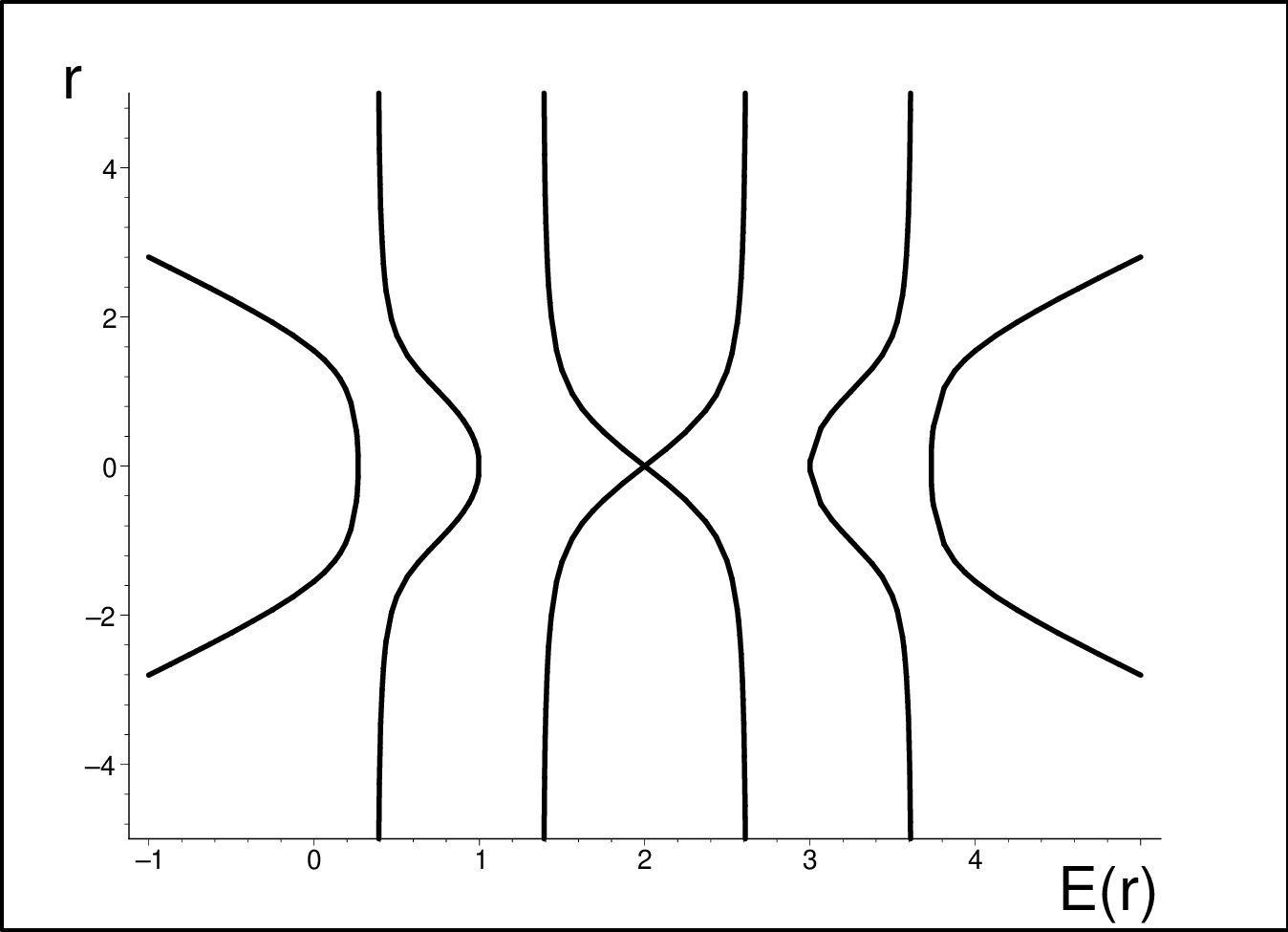,angle=270,width=0.5\textwidth}
\end{center}
\vspace{-2mm}\caption{Graphical version of the closed-form
representation (\ref{17}) of the strictly real
bound-state energy spectrum
of Schr\"{o}dinger Eq.~(\ref{Ka8p}) at $N=6$
with specific $z=z(r)={\rm i} \sqrt{1-r^2}$.
 \label{figone}}
\end{figure}

The methodical message delivered by the latter analysis can easily
be extended to any $N$.
For example, once we choose
$N=6$,
the evaluation
{of}
energies $E_n^{(6)}$
seems to be a purely numerical task because the secular
{determinant} $\det (H-E)$
{is equal to polynomial}
 $$
 {E}^{6}-12\,{E}^{5}+ \left( 56-{r}^{2} \right)
E^{4}+ \left( -128+8\,{r}^{2} \right) {E}^{3}+ \left(
147-21\,{ r}^{2} \right) {E}^{2}+ \left( -76+20\,{r}^{2}
\right) E +12-5\,{r}^{2}
 $$
of the sixth degree in $E$.
Fortunately, {this polynomial}
is just a linear function of $r^2$. Thus,
the $N=6$ spectrum can {\em exactly\,} be defined by
the following closed implicit-function formula
 \be
r_\pm=r_\pm(E)=
  \pm \sqrt {
 \frac
   { {E}^{4}-8\,
{E}^{3}+ 20\,{E}^{2}-16\,E+3 }
     {{E}^{4}-8\,{E}^{3}+21\,{E}^{2}-20\,E+5 }
 }
\left(
E-2 \right)
\,.
\label{17}
 \ee
This is an analytic result which could be extended to any finite
Hilbert-space dimension $N$. Numerically it is complemented
by Figure \ref{figone} where we
can see that the EP singularity
(also known as ``non-Hermitian degeneracy'' \cite{Berry})
at $r=0$ manifests itself
by the merger of two levels in the middle of the spectrum.

\section{Non-stationary inner-product metrics\label{s4I}}

In our present toy model the nontrivial (i.e., non-Hermitian {\em
and\,} time-dependent) dynamics
{is}
introduced via boundary
conditions. {Partially, it can simplify}
a consequent application
of the {theory}.

\subsection{Definition}

Once we managed to confirm the reality
{of}
spectrum we have to
move to the next model-building task which lies, in both of the NSP
and NIP contexts, in the construction or selection of the physical
inner product metric $\Theta$ which would be compatible
{with}
quasi-Hermiticity constraint (\ref{tJD}). At this stage of
development we may feel encouraged by the observation that for the
same but stationary, time-independent interaction (i.e., in the
simpler NSP dynamical regime), incidentally, a closed-form solution
$\Theta(H)$ of Eq.~(\ref{tJD}) appeared available
\cite{Krejcirik}.

One of the most universal construction strategies which might be
also used in the NIP setting has been described in \cite{SIGMAdva}.
After a restriction of attention to the unitary quantum models
living in the finite-dimensional Hilbert spaces we showed there that
whenever one manages to solve the conjugate version
{of}
Schr\"{o}dinger eigenvector problem or, in the notation of
Ref.~\cite{SIGMA}, of the ketket-vector problem
 \be
 H^\dagger_{}(t)\,|\xi_{n}(t)\kkt
 =E_n\,|\xi_{n}(t)\kkt\,,
 \ \ \ \ n=1,2,\ldots,N\,
 \label{etwoSE}
 \ee
then {\em all\,} of the eligible (i.e., invertible and positive
definite \cite{Geyer})
inner-product metrics form an $N-$parametric
family,
 \be
 \Theta_{(\vec{\kappa}(t))}(t)=
 \sum_{n}\,|\xi_{n}(t)\kkt\,
 \kappa_n^{}(t)\,\bbr  \xi_{n}(t)|\,,
 \ \ \ \
 \forall\ \kappa_n^{}(t)>0
 \,.
 \label{onemet}
 \ee
Different physics becomes represented by the different choices
{of}
parameters {$\kappa_n^{}(t)>0$}. In what follows, for the sake
of simplicity, we will
work just with the trivial choice of $\kappa_n^{(constant)}(t)=1$
and we will speak about a ``special'' {\it alias\,} ``standard'' choice of
$\Theta_{(\vec{1})}(t)=\Theta_S(t)$.

\subsection{Non-stationary Dyson maps }

Let us now fully concentrate on the non-stationary
NIP model-building dynamical scenarios with the
Hilbert-space metrics which vary with time.
Keeping in mind that this time-dependence is essential,
contributing to an enormous increase of the model-building difficulties
caused by the transition from the NSP formalism to its NIP generalization.

The first {duty} to {fulfill}
is {the}
Dyson-map factorization of the
manifestly non-statinary metric $\Theta(t)=\Omega^\dagger(t)\Omega(t)$.
Often, this is performed via
taking a self-adjoint square root of $\Theta(t)$
followed by a completion of the construction by considering also
the unitary-matrix ambiguity
{of}
eligible $\Omega(t)$s.
As long as such a
{task} seems enormously complicated,
its upgrade will be based here on a return to
{a {\em diagonal\,}}
Hamiltonian matrix $\mathfrak{h}_S(t)$
(cf. its $N=2$ sample (\ref{17S})).
Within such a project, definition (\ref{etwoSE})
of the ketkets has to be re-read
as
{a}
matrix intertwining problem
 \be
 H^\dagger(t)\,\Omega_S^\dagger(t)=\Omega_S^\dagger(t)\,\mathfrak{h}_S(t)\,
 \label{emonemet}
 \ee
where the matrix of eigenvalues $\mathfrak{h}_S(t)$ is diagonal.

Relation (\ref{emonemet}) can be interpreted
as a specific realization
{of}
correspondence between
Hilbert
spaces {${\cal L}\leftrightarrow {\cal H}_{friendly}$}
as defined in
Eq.~(\ref{[7]}).
In this sense the set of all of
the ketket eigenvectors $|\xi_{n}(t)\kkt$
of $H^\dagger(t)$ (cf. Eq.~(\ref{etwoSE}))
can be re-visualized as the set of separate
columns
{of}
matrix $\Omega_S^\dagger(t)$.
As a consequence, we can put
 \be
 \Theta_{S}(t)=\Omega_S^\dagger(t)\,\Omega_S(t)\,
 \label{monemet}
 \ee
and speak about the metric in which
the change
{of}
normalization
of the eigenketkets in (\ref{etwoSE})
(which is admissible)
can be perceived as
equivalent to the change of
parameters $\vec{\kappa}(t)$
in (\ref{onemet}).

\section{Non-numerical results at $N=2$\label{sIIIc2}}

Whenever we are given the non-stationary and non-Hermitian
operator $H(t)$ representing the
instantaneous real and observable bound-state energies,
the main obstacle on our way towards the
{tests}
(i.e., towards the
predictions represented by formula (\ref{eme}))
is the necessity
{of}
construction of the generator
$G(t)=H(t)-\Sigma(t)$
{of}
evolution of the
relevant ket vectors (cf. Eq. (\ref{DCauchy})).

In the literature, very often, the authors circumvent the problem
and, typically, complement the knowledge of $H(t)$
by some additional information about $G(t)$ or $\Sigma(t)$.
In our present paper we intend to argue that the dynamical input
knowledge of $H(t)$ can be, in some cases and sense, sufficient.

The latter statement {is}
strongly model-dependent.
Even in the case of our present
specific and sufficiently elementary discrete square-well model (\ref{Ka8t}),
an
explicit and exhaustive description of
its properties would be, in spite of its solvability, complicated and not too
illuminative, especially at the larger Hilbert-space dimensions $N$.
{As}
long as we intend to provide here
just an overall qualitative support of
the user-friendliness
{of}
applicability of the NIP-based models,
we will restrict
our
{attention}
to the
{mere}
first nontrivial
version
{(\ref{prodva}) of our}
model with $N=2$.

Surprisingly enough,
the results of this study will be shown to be not only encouraging
but also compact and, in a way, persuasive and sufficiently informative.

\subsection{The first task: closed-form Dyson map}

During our preliminary search for the closed-form solutions of the
$N=2$ matrix version of Schr\"{o}dinger Eq.~(\ref{etwoSE})
{\it alias\,} (\ref{emonemet})
we tried to use several computer-assisted symbolic-manipulation
techniques, and we failed.
The success only came with a return to the paper-and-pencil techniques.
In the model of Eq.~(\ref{prodva}) they
guided us to perform another change
of variables
setting $r=\sin \varphi$ where
$\varphi=\varphi(t) \neq \varphi^{(EP)} =  0, \pm \pi, \ldots$.

{We}
will drop, in some cases, the ubiquitous
time-dependence-emphasizing brackets $(t)$ as redundant.
In particular, having turned attention
{to}
non-stationary
conjugate-Hamiltonian operator
  $$
 H^\dagger=
 \left[ \begin {array}{cc} 2+i\cos \varphi&-1\\\noalign{\medskip}
-1&2-i\cos \varphi\end {array} \right]
 $$
we got not only the above-mentioned spectrum $E_\pm^{(2)}= 2+\sin \varphi$
(where one can simulate $\pm$ by ${\rm sign\,} \varphi$
and omit the subscript) but also,
having solved Eq.~(\ref{emonemet}), one of the most compact eligible
{non-stationary}
forms
of the {respective}
conjugate and non-conjugate Dyson maps,
  $$
 \Omega_S^\dagger=
 \left[ \begin {array}{cc}
 1&-{\rm i}\,\exp {\rm i}\varphi
 \\\noalign{\medskip}
 {\rm i}\,\exp {\rm i}\varphi &1
 \end {array} \right]\,,\ \ \ \
 \Omega_S=
 \left[ \begin {array}{cc}
 1&-{\rm i}\,\exp (-{\rm i}\varphi)
 \\\noalign{\medskip}
 {\rm i}\,\exp (-{\rm i}\varphi) &1
 \end {array} \right]\,
  $$
Immediately, this yields the metric,
 \be
 \Theta_{S}=\Omega_S^\dagger\,\Omega_S=
  \left[ \begin {array}{cc}
 2&-2{\rm i}\,\cos \varphi
 \\\noalign{\medskip}
2 {\rm i}\,\cos \varphi &2
 \end {array} \right]
 \,.
 \label{umonemet}
 \ee
This formula coincides with the one constructed in \cite{2014}
where we, unfortunately, did not find the way towards
its Dyson-map factorization. A consistency of the latter metric
proves also supported, off the EP singularity, by the
positivity of its two eigenvalues
$\theta_\pm = 2 \pm 2\,\cos \varphi$
as well as
by its
diagonality
and proportionality to a unit operator
in the Hermitian-Hamiltonian limit of $\cos \varphi \to 0$.

\subsection{The second task: closed-form Coriolis force}

By the direct computation we
{get}
matrices
  $$
 \Omega_S^{-1}=\frac{1}{1-\exp (-2{\rm i}\varphi)}\,
 \left[ \begin {array}{cc}
 1&{\rm i}\,\exp (-{\rm i}\varphi)
 \\\noalign{\medskip}
 -{\rm i}\,\exp (-{\rm i}\varphi) &1
 \end {array} \right]\,,\ \ \ \ \
 \dot{\Omega}_S=\dot{\varphi}\cdot
 \left[ \begin {array}{cc}
 0&-\exp (-{\rm i}\varphi)
 \\\noalign{\medskip}
 \exp (-{\rm i}\varphi) &0
 \end {array} \right]\,
  $$
as well
{as}
Coriolis force,
  \be
 \Sigma_S
 ={\rm i} \Omega_S^{-1}\dot{\Omega}_S
 =
 \frac{\dot{\varphi}}{2\,\sin \varphi}\,
 \left[ \begin {array}{cc}
 {\rm i}\exp (-{\rm i}\varphi)&-1
 \\\noalign{\medskip}
 1 &{\rm i}\exp (-{\rm i}\varphi)
 \end {array} \right]
 \,.
 \label{tenhle}
  \ee
This result leads us to an
important observation
that {\em both\,} of the eigenvalues $\sigma_\pm$ of $\Sigma_S$
are {\em always\,} complex,
 $$
 \sigma_\pm=\left (
 1+{\rm i}\,\frac{\cos \varphi \pm 1}{\sin \varphi}
 \right )\cdot {\dot{\varphi}}/{2}
 =\left \{
 \ba
  \left (
 1+{\rm i}\,{\cot \varphi/2}
 \right ){\dot{\varphi}}/2\\
 \left (
 1-{\rm i}\,{\tan \varphi/2}
 \right ){\dot{\varphi}}/2\,.
 \ea\right .
 $$
Moreover, these two eigenvalues do not even form a complex conjugate doublet.

This means that there exists no operator of parity ${\cal P}$ which could make the
Coriolis force (i.e., the Heisenberg-equation generator
{\it alias\,} ``Heisenberg Hamiltonian'') ${\cal PT}-$symmetric.

\subsection{The ultimate task: closed-form Schr\"{o}dinger equation}

Once we abbreviate $D={\dot{\varphi}(t)}/({2\,\sin \varphi (t)})$ and
set $1-D=A=A(t)$ and $1+D=B=B(t)$, we may
recall Schr\"{o}dinger
Eq.~(\ref{DCauchy})
and evaluate
the difference $G_S(t)=H(t)-\Sigma_S(t)${,}
 $$
 G_S =
  \left[ \begin {array}{cc}
  2-D\sin \varphi- {\rm i} B \cos \varphi
  &-A
  \\\noalign{\medskip}
 -B
 &
 2-D\sin \varphi+ {\rm i} A \cos \varphi
 \end {array} \right]\,.
 $$
Its eigenvalues are available in compact form $g_\pm= 2-D\sin \varphi+w_\pm$
with
 $$
 w_\pm = - {\rm i}D\cos \varphi \pm \sqrt{\sin^2 \varphi - D^2}\ .
 $$
In the ``almost stationary'' dynamical regime with
small ${\dot{\varphi}}$ such that
$D^2 < \sin^2 \varphi$ we get
 $$
 w_\pm
 %=  \pm ({\sin^2 \varphi - D^2})^{1/2}- {\rm i}D\cos \varphi
 %=  \pm \sin \varphi \,({1 - D^2/\sin^2 \varphi})^{1/2}- {\rm i}D\cos \varphi
 =\pm \sin \varphi- {\rm i}D\cos \varphi + {\cal O}(D^2)
 $$
yielding the two strictly non-real eigenvalues of $G_S$ which are even not mutually conjugate,
$$
g_\pm= 2 \pm \sin \varphi-D\sin \varphi- {\rm i}D\cos \varphi + {\cal O}(D^2)\,.
$$
In the opposite, ``strongly non-stationary'' case with
small $ \sin^2 \varphi < D^2 $ we get
   $$
 w_\pm
 =  {\rm i} D
 \left [- \cos \varphi\pm
 ({1- D^{-2}\sin^2 \varphi })^{1/2}
 \right ]\,
 %=  \pm \sin \varphi \,({1 - D^2/\sin^2 \varphi})^{1/2}- {\rm i}D\cos \varphi
 %=\pm \sin \varphi- {\rm i}D\cos \varphi + {\cal O}(D^2)
 $$
so that
the whole correction
reflecting the influence of the Coriolis force (i.e., of
{operator (\ref{tenhle})
proportional to} ${\dot{\varphi}}\neq 0$)
becomes strictly purely imaginary.

Again,
the two strictly non-real eigenvalues of $G_S$ are not mutually conjugate.
This implies that such a {``Schr\"{o}dinger Hamiltonian''
{\it alias\,}}
generator of evolution
{of}
state vectors
can never be required ${\cal PT}-$symmetric.

\section{Discussion\label{sIVdi}}

In the overall context
of our present paper it is worth inserting a
terminological remark that
the widely used and popular word ``non-Hermitian'' could be
potentially
misleading
and deserves explanation: What
the majority
{of}
authors of the reviews and papers
on the subject
had in mind was a
mathematical formalism
which just
describes the conventional
unitary {\it alias\,} closed
quantum systems in an innovative NSP or NIP representation.

\subsection{Square-well quantum models\label{ewsIr}}

The idea of the
more or less revolutionary
replacement
{of}
conventional self-adjoint Hamiltonians (say, $\mathfrak{h}$)
by their
less usual non-Hermitian isospectral
but, presumably, user-friendlier
partners $H = \Omega^{-1}\,\mathfrak{h}\,\Omega$
can be
traced back
{to}
Dyson's paper \cite{Dyson}.
He introduced the notion
{of an}
invertible
preconditioning operator $\Omega$ which has been allowed
stationary but non-unitary. This made
his followers able to
define a Hamiltonian-dependent
inner-product Hilbert-space metric \cite{Geyer},
the knowledge of which enabled them to re-read the
self-adjointness of $\mathfrak{h}=\mathfrak{h}^\dagger$,
formally at least,
as equivalent to the
{NSP (and, later, also to the NIP)}
quasi-Hermiticity of $H$.

In the related innovative model-building process, the simplicity
{of}
non-Hermitian $H$ with real spectrum was essential.
This was the reason why
some of the most user-friendly
quasi-Hermitian generalizations of models (\ref{spojhambc})
and/or (\ref{diskrhambc}) were only modified
``minimally'', by the mere change of
boundary condition. In this setting, the discrete-square-well
dynamics controlled by certain
non-Hermitian but PT-symmetric and {\em time-independent,
stationary\,} boundary conditions can be found described in our
older {NSP} paper \cite{2014}.

Among the methodically welcome features of
this {(i.e., still just stationary)}
model we may mention its exact solvability. In certain
intervals of parameters the bound-state energies were
real and given as roots of certain elementary
trigonometric expressions. Also the wave-functions were
expressed, in {\it loc. cit.},
{in}
closed form.
The model has
been rendered {quasi-Hermitian}
by means of an
explicit construction of a nontrivial {NSP}
inner-product metric
{$\Theta$.}

From the purely methodical point of view the assumption
{of}
stationarity of the model was essential because it enabled us to
recall and apply just the formulation of quantum mechanics of
reviews \cite{Geyer,ali,Carl}. In this context we were able to
reduce the analysis to the mere diagonalization of a
non-Hermitian $N$ by $N$ matrix.
The role and influence
{of}
boundary conditions were represented
by the single complex time-independent parameter.
{In this sense the present, NIP-based
non-stationary
extension of the model
of paper \cite{2014}
can be perceived as opening broad new horizons.}

\subsection{The problem of observables}

In the majority of publications
{on}
hiddenly Hermitian
models
{the authors are accepting}
the assumption
{of}
stationarity of
the
Dyson's
{map in Eq.~(\ref{[7]}). For several
good reasons:
One of the most important ones is purely technical because
the assumption of stationarity}
implies the full
formal equivalence between ``the old'' Schr\"{o}dinger equation (\ref{pCauchy})
and ``the new'' NSP Schr\"{o}dinger equation (\ref{SCauchy}).
The self-adjointness of $\mathfrak{h}$ can be perceived as
equivalent to the quasi-Hermiticity of $H$.
Thus, given a stationary non-Hermitian $H$ with real spectrum,
a key to the
completion of the
{NSP}
theory can be seen
in the specification of
{such}
a
self-adjoint and positive definite operator $\Theta$ which would
make the operators of
{such NSP observables}
quasi-Hermitian \cite{Geyer}.

A weak point of such a philosophy lies in the
necessity
{of}
formulation of the dynamical-input information in
terms of the operators of observables which are non-Hermitian.
Usually, one succeeds in choosing a sufficiently interesting
non-Hermitian candidate $H$ for the energy-representing NSP
Hamiltonian, say, in its Klein-Gordon form \cite{aliKG}, or in its
Proca-field version \cite{Smejkal,Smejkalb}, etc. Nevertheless, the
resulting non-triviality
{of}
physical metric $\Theta\neq I$
becomes a source of difficulties. It implies that any other eligible
observable (say, $\Lambda$) must satisfy the
hidden-Hermiticity relation with {\em the same\,} metric
$\Theta=\Theta(H)$,
 \be
 \Lambda^\dagger\,\Theta(H)=\Theta(H)\,\Lambda\,.
 \label{lamJD}
 \ee
{In this light it is obvious that
in the future studies, more attention will have to be paid to}
the most fundamental concept of
the observable spatial coordinate $\Lambda_x$.
{Indeed, the task of its construction}
becomes highly nontrivial even for the
{stationary}
square-well
potentials \cite{Batal} or for various
{even more elementary}
delta-function interactions
\cite{Jones}.
{Naturally, also in such a
context the present, NIP-based
extension of the model-building philosophy
to the
non-stationary-metric domain
opens new methodical as well as phenomenological
challenges and questions.
{\it Pars pro toto,}
what
becomes particularly important
is
the role of the
exact solvability
as sampled by our illustrative model, and
as rendered possible by its simplified,
boundary-controlled dynamics.}

\section{Summary\label{sIr}}

{In our present paper, non-stationary version of
unitary quantum mechanics formulated in
non-Hermitian (or, more precisely, in hiddenly Hermitian)
interaction-picture representation was recalled and
illustrated. The purpose was served by an
elementary $N$ by $N$ matrix Hamiltonian $H(t)$ mimicking a 1D-box
system in which the physics is controlled by time-dependent boundary
conditions.}

{The model was presented as analytically solvable at
$N=2$. {\it Expressis verbis} this means that for both of the
underlying Heisenberg and Schr\"{o}dinger evolution equations the
generators (i.e., in our notation, the respective operators
$\Sigma(t)$ and $G(t)$) became available in closed form. In this light,
the key
message delivered by our paper
is that contrary to the conventional beliefs and in spite of
the unitarity
{of}
evolution of the system, neither its
``Heisenbergian Hamiltonian'' $\Sigma(t)$ nor its
``Schr\"{o}dingerian Hamiltonian'' $G(t)$ possesses a real
spectrum (or even some spectrum containing the conjugate pairs of
complex eigenvalues).}

{Such an observation can be perceived as
being of paramount importance}
in the quickly developing field of study of the role of
non-self-adjoint operators in quantum physics.
{One} of the technically
most relevant division lines separates,
{in this context,} the stationary from
non-stationary models \cite{book}. This observation motivated also
our present paper. We imagined that at least some of the existing
solvable stationary models still wait for a non-stationary
extension.

In {the older}
review paper \cite{ali} we read that
the inner product metric ``cannot depend on time, unless \ldots
[operator $G(t)$] is not observable.'' In fact, the latter
non-observability is easy to accept and, after all, fully compatible
with our present results as well as with the explicit theoretical
description
{of}
unitary quantum dynamics.
{Indeed, one can work, in a fully consistent manner,
with a broad class of}
non-observable operators $G(t)$ (cf. \cite{timedep} and also a few
later confirmations and reconfirmations of this observation in
papers \cite{FrFrith,Tara,Android}).

{Naturally,}
the transition
{to}
non-Hermitian {\em and\,} time-dependent operators of observables
{leads to}
multiple {new - and not always} expected -
technical obstacles.
{This is, in fact,} the
main weakness of
{our present}
non-stationary amendment
{of the more common}
stationary
{models.}
For this reason the early attention of
researchers turned to the {non-stationary models
which were} exactly solvable
\cite{SIGMA,Bila,Bilab,ITJPb}. Only recently, the progress
{in our}
understanding of various technical subtleties led to the more
systematic analyses
{and to the}
less schematic methodical
considerations \cite{Fring,ITJP}. Still, the exactly solvable models
keep playing
{a}
dominant role.

In parallel,
{suitable}
approximative techniques had to be
developed \cite{Maamache,Rebeka,Rebekab}. Several new directions of
applicability of the NIP constructive philosophy {and} of its
innovative modifications emerged \cite{Frith,Bishop,WDW}. Among the
most recent ones let {we would like to}
recall paper \cite{Fermi} by
Fring and Taira in which the authors were able to study the
time-dependent boundary conditions in an implementation to the well
known Swanson's ``benchmark'' non-Hermitian Hamiltonian
\cite{Swanson,Swansonb}.

%\newpage

\end{document}